\documentclass[aps,preprint]{revtex4}
\usepackage{amssymb,epsf}
\usepackage{amsmath}
\usepackage{graphicx}

\begin{document}
\title{Thermodynamics of Rotating Black Branes in
Gauss-Bonnet-Born-Infeld Gravity}
\author{M. H. Dehghani$^{1,2}$\footnote{email address:
mhd@shirazu.ac.ir} and S. H. Hendi$^{1,3}$\footnote{email address:
hendi@mail.yu.ac.ir}}

\affiliation{$^1$Physics Department and Biruni Observatory,
College of Sciences, Shiraz
University, Shiraz 71454, Iran \\
$^2$Research Institute for Astrophysics and Astronomy of Maragha
(RIAAM), Maragha, Iran\\
$^3$ Physics Department, Yasouj University, Yasouj, Iran}

\begin{abstract}
Considering both the Gauss-Bonnet and the Born-Infeld terms, which
are on similar footing with regard to string corrections on the
gravity side and electrodynamic side, we present a new class of
rotating solutions in Gauss-Bonnet gravity with $k$ rotation
parameters in the presence of a nonlinear electromagnetic field.
These solutions, which are asymptotically anti-de Sitter in the
presence of cosmological constant, may be interpreted as black
brane solutions with inner and outer event horizons, an extreme
black brane or naked singularity provided the metric parameters
are chosen suitably. We calculate the finite action and conserved
quantities of the solutions by using the counterterm method, and
find that these quantities do not depend on the Gauss-Bonnet
parameter. We also compute the temperature, the angular
velocities, the electric charge and the electric potential. Then,
we calculate the entropy of the black brane through the use of
Gibbs-Duhem relation and show that it obeys the area law of
entropy. We obtain a Smarr-type formula for the mass as a function
of the entropy, the angular momenta and the charge, and show that
the conserved and thermodynamic quantities satisfy the first law
of thermodynamics. Finally, we perform a stability analysis in
both the canonical and grand-canonical ensemble and show that the
presence of a nonlinear electromagnetic field has no effect on the
stability of the black branes, and they are stable in the whole
phase space.
\end{abstract}

\maketitle

\section{Introduction}

In recent years a renewed interest in the Lovelock gravity and
Born-Infeld electrodynamics has been appeared. This is due to the
fact that both of them emerge in the low energy limit of string
theory \cite{frad,berg,met}. On the gravity side, string theories
in their low-energy limit give rise to effective models of gravity
in higher dimensions which involves higher curvature terms, while
on the electrodynamics side the Born-Infeld action has been
occurring with the development of superstring theory, where the
dynamics of D-branes and some soliton solutions of supergravity is
governed by the Born-Infeld action. While the Lovelock gravity was
proposed to have field equations with at most second order
derivatives of the metric \cite{Lov}, the nonlinear
electrodynamics proposed, by Born and Infeld, with the aim of
obtaining a finite value for the self-energy of a point-like
charge~ \cite{BI}. The Lovelock gravity reduces to Einstein
gravity in four dimensions and also in the weak field limit, while
the Lagrangian of the Born-Infeld (BI) electrodynamics reduces to
the Maxwell Lagrangian in the weak field limit. There has been
considerable works on both of these theories. In Lovelock gravity,
there has been some attempts for understanding the role of the
higher curvature terms from various points of view, especially
with regard to black hole physics. For example, exact static
spherically symmetric black hole solutions of the second order
Lovelock (Gauss-Bonnet) gravity have been found in Ref.
\cite{Des}, and of the Gauss-Bonnet-Maxwell model in Ref.
\cite{Wil1}. The thermodynamics of the uncharged static
spherically black hole solutions has been considered in \cite{MS},
of solutions with nontrivial topology in \cite{Cai} and of charged
solutions in \cite{Wil11,Od11}. All of these known solutions in
Gauss-Bonnet gravity are static. Not long ago one of us introduced
two new classes of rotating solutions of second order Lovelock
gravity and investigated their thermodynamics \cite{Deh1, Deh2},
made the first attempt for finding exact static solutions in third
order Lovelock gravity with the quartic terms \cite{DSh}, and
presented the charged rotating black brane solutions of third
order Lovelock gravity \cite{DM1}. Recently, the Taub-NUT/bolt
solutions of Gauss-Bonnet and Gauss-Bonnet-Maxwell gravity have
been constructed \cite{DM2,Hendi}. The first aim to relate the
nonlinear electrodynamics and gravity has been done by Hoffmann
\cite {Hoffmann}. He obtained a solution of the Einstein equations
for a point-like Born-Infeld charge, which is devoid of the
divergence of the metric at the origin that characterizes the
Reissner-Nordstr\"{o}m solution. However, a conical singularity
remained there, as it was later objected by Einstein and Rosen.
The spherically symmetric solutions in Einstein-Born-Infeld (EBI)
gravity with or without a cosmological constant have been
considered by many authors \cite{EBI,Dey}. Recently, the rotating
black brane solutions of Einstein-Born-Infeld gravity have been
constructed, and their thermodynamics has been
investigated.\cite{Rastegar}.

Considering the analogy between the Gauss-Bonnet and the
Born-Infeld terms, which are on similar footing with regard to
string corrections on the gravity side and electrodynamic side,
respectively, it is plausible to include both these corrections
simultaneously. Indeed, if one is to consider the Maxwell fields
coupled to a gravitational action, which also includes string
generated corrections at higher orders, then, it is naturally
important to consider string generated corrections to the
electromagnetic field action as well. Such corrections come from a
coupling of abelian gauge fields to open bosonic or open
superstrings. To be precise, it is known that, just like the
Gauss-Bonnet terms, there are Born-Infeld terms which appear as
higher order corrections to the Maxwell action~\cite{frad}. The
static black hole solutions of Gauss-Bonnet-Born-Infeld (GBBI)
gravity have been constructed in Ref. \cite{Wil2}. In this paper,
we attempt to construct the rotating black brane solutions of GBBI
gravity and investigate their properties. We also perform a
stability analysis and investigate the effects of nonlinear BI
field on the stability of the solutions. The rest of the paper is
organized as follows. In section II, we write down the first three
term of Lovelock gravity (Gauss-Bonnet) in $(n+1)$-dimensions with
Born-Infeld type corrections of Maxwell equations. In section III,
we consider a new class of rotating black brane and solve it for
gravitation and electromagnetic field equations. We further
discuss their thermodynamic properties in section IV. Section V
has include of stability of black brane in canonical and
grand-canonical ensemble. We finish our paper with concluding
remarks in section IV.

\section{Field Equations in Gauss-Bonnet-Born-Infeld Gravity}

\label{Fiel} \label{Pot1b} The most fundamental assumption in standard
general relativity are the requirement of general covariance and that the
field equations be second order. Based on these principles, the most general
Lagrangian in arbitrary dimensions is the Lovelock Lagrangian. The
gravitational action of Lovelock theory can be written as \cite{Lov}
\begin{equation}
I_{G}=\int d^{n+1}x\sqrt{-g}\sum_{i=0}^{[(n+1)/2]}\alpha _{i}\mathcal{L}_{i}
\label{Lov1}
\end{equation}
where $[z]$ denotes the integer part of $z$, $\alpha _{i}$ is an arbitrary
constant and $\mathcal{L}_{i}$ is the Euler density of a $2i$-dimensional
manifold,
\begin{equation}
\mathcal{L}_{i}=\frac{1}{2^{i}}\delta _{\rho _{1}\sigma _{1}\cdots \rho
_{J}\sigma _{J}}^{\mu _{1}\nu _{1}\cdots \mu _{J}\nu _{J}}R_{\mu _{1}\nu
_{1}}^{\phantom{\mu_1\nu_1}{\rho_1\sigma_1}}\cdots R_{\mu _{J}\nu _{J}}^{%
\phantom{\mu_J \nu_J}{\rho_J \sigma_J}}  \label{Lov2}
\end{equation}
In Eq. (\ref{Lov2}), $\delta _{\rho _{1}\sigma _{1}\cdots \rho _{J}\sigma
_{J}}^{\mu _{1}\nu _{1}\cdots \mu _{J}\nu _{J}}$ is the generalized totally
anti-symmetric Kronecker delta and $R_{\mu \nu }^{\phantom{\mu\nu}{\rho%
\sigma}}$ is the Riemann tensor. We note that in $(n+1)$-dimensions, all
terms for which $d>[(n+1)/2]$ are identically equal to zero, where $[x]$
denotes the integer part of $x$, and the term $d=(n+1)/2$ is a topological
term. Consequently only terms for which $d<(n+1)/2$ contribute to the field
equations. Here we study Gauss-Bonnet gravity, that is first three terms of
Lovelock gravity. In this case the action of GBBI gravity is
\begin{eqnarray}
I_{G} &=&-\frac{1}{16\pi }\int_{\mathcal{M}}d^{n+1}x\sqrt{-g}\left\{
R-2\Lambda +\alpha (R_{\mu \nu \gamma \delta }R^{\mu \nu \gamma \delta
}-4R_{\mu \nu }R^{\mu \nu }+R^{2})+L(F)\right\}  \nonumber \\
&&-\frac{1}{8\pi }\int_{\partial \mathcal{M}}d^{n}x\sqrt{-\gamma
}\left\{ \Theta +2\alpha \left( J-2\widehat{G}_{ab}\Theta
^{ab}\right) \right\} \label{IG}
\end{eqnarray}
where $\Lambda =-n(n-1)/2l^{2}$ is the cosmological constant, $\alpha $ is
the Gauss-Bonnet coefficient with dimension $(\mathrm{length})^{2}$, $R$, $%
R_{\mu \nu }$ and $R_{\mu \nu \gamma \delta }$\ are the Ricci scalar and
Ricci and Riemann tensors of the manifold $\mathcal{M}$, $L(F)$ is the
Lagrangian of Born-Infeld
\begin{equation}
L(F)=4\beta ^{2}\left( 1-\sqrt{1+\frac{F^{2}}{2\beta ^{2}}}\right) ,
\label{ActEB}
\end{equation}
In Eq. (\ref{ActEB}), $\beta $\ is called the Born-Infeld parameter with
dimension of mass, $F^{2}=F^{\mu \nu }F_{\mu \nu }$\ where $F_{\mu \nu
}=\partial _{\mu }A_{\nu }-\partial _{\nu }A_{\mu }$\ is electromagnetic
tensor field and $A_{\mu }$\ is the vector potential\textbf{.} In the limit $%
\beta \rightarrow \infty $, $L(F)$ reduces to the standard Maxwell form $%
L(F)=-F^{2}$, while $L(F)\rightarrow 0$ as $\beta \rightarrow 0$. The second
integral in Eq. (\ref{IG}) is a boundary term which is chosen such that the
variational principle is well defined \cite{MyeDavis}. In this term, $\gamma
_{ab}$ is induced metric on the boundary $\partial \mathcal{M}$, $\Theta $
is trace of extrinsic curvature $\Theta ^{ab}$ of this boundary, $\widehat{G}%
^{ab}(\gamma )$ is Einstein tensor calculated on the boundary, and $J$ is
trace of:
\begin{equation}
J_{ab}=\frac{1}{3}(\Theta _{cd}\Theta ^{cd}\Theta _{ab}+2\Theta \Theta
_{ac}\Theta _{b}^{c}-2\Theta _{ac}\Theta ^{cd}\Theta _{db}-\Theta ^{2}\Theta
_{ab})  \label{psi}
\end{equation}

Varying the action with respect to the metric tensor $g_{\mu \nu }$ and
electromagnetic field $A_{\mu }$ the equations of gravitation and
electromagnetic fields are obtained as
\begin{eqnarray}
&&G_{\mu \nu }+\Lambda g_{\mu \nu }-\alpha \{4R^{\rho \sigma }R_{\mu \rho
\nu \sigma }-2R_{\mu }^{\ \rho \sigma \lambda }R_{\nu \rho \sigma \lambda
}-2RR_{\mu \nu }+4R_{\mu \lambda }R_{\text{ \ }\nu }^{\lambda }  \nonumber \\
&&+\frac{1}{2}g_{\mu \nu }(R_{\kappa \lambda \rho \sigma }R^{\kappa \lambda
\rho \sigma }-4R_{\rho \sigma }R^{\rho \sigma }+R^{2})\}=\frac{1}{2}g_{\mu
\nu }L(F)+\frac{2F_{\mu \lambda }F_{\phantom{\lambda}{\mu}}^{\lambda }}{%
\sqrt{1+\frac{F^{2}}{2\beta ^{2}}}},  \label{Geq}
\end{eqnarray}

\begin{equation}
\partial _{\mu }\left( \frac{\sqrt{-g}F^{\mu \nu }}{\sqrt{1+\frac{F^{2}}{%
2\beta ^{2}}}}\right) =0,  \label{BIeq}
\end{equation}
where $G_{\mu \nu }$ is the Einstein tensor. Equation (\ref{Geq}) does not
contain the derivative of the curvatures, and therefore the derivatives of
the metric higher than two do not appear. Thus, the Gauss-Bonnet gravity is
a special case of higher derivative gravity.

In general the action $I_{G}$, is divergent when evaluated on the solutions,
as is the Hamiltonian and other associated conserved quantities. A
systematic method of dealing with this divergence in Einstein gravity is
through the use of the counterterms method inspired by the anti-de Sitter
conformal field theory (AdS/CFT) correspondence \cite{Mal}. This conjecture,
which relates the low energy limit of string theory in asymptotically anti
de-Sitter spacetime and the quantum field theory living on the boundary of
it, have attracted a great deal of attention in recent years. This
equivalence between the two formulations means that, at least in principle,
one can obtain complete information on one side of the duality by performing
computation on the other side. A dictionary translating between different
quantities in the bulk gravity theory and their counterparts on the boundary
has emerged, including the partition functions of both theories. This
conjecture is now a fundamental concept that furnishes a means for
calculating the action and conserved quantities intrinsically without
reliance on any reference spacetime \cite{Sken1BKOd1}. It has also been
applied to the case of black holes with constant negative or zero curvature
horizons \cite{Deh3} and rotating higher genus black branes \cite{Deh4}.
Although the AdS/CFT correspondence applies for the case of a specially
infinite boundary, it was also employed for the computation of the conserved
and thermodynamic quantities in the case of a finite boundary \cite{Deh5}.

All of the work mention in the last paragraph was limited to Einstein
gravity. Although the counterterms in Lovelock gravity should be a scalar
constructed from Riemann tensor as in the case of Einstein gravity, they are
not known for the case of Lovelock gravity till now. But, for the solutions
with flat boundary, $\widehat{R}_{abcd}(\gamma )=0$, there exists only one
boundary counterterm
\begin{equation}
I_{\mathrm{ct}}=-\frac{1}{8\pi }\int_{\partial \mathcal{M}}d^{n}x\sqrt{%
-\gamma }\left( \frac{n-1}{l_{\mathrm{eff}}}\right) ,  \label{Ict}
\end{equation}
where $l_{\mathrm{eff}}$\ is a scale length factor that depends on $l$ and $%
\alpha $, that must reduce to $l$ as $\alpha $ goes to zero. One may note
that this counterterm has exactly the same form as the counterterm in
Einstein gravity for a spacetime with zero curvature boundary in which $l$
is replaced by $l_{\mathrm{eff}}$. The total action can be written as a
linear combination of the action of bulk, boundary (\ref{IG}) and the
counterterm (\ref{Ict})
\begin{equation}
I=I_{G}+I_{\mathrm{ct}}  \label{Itot}
\end{equation}
Having the total finite action, one can use the Brown-York definition of
stress energy momentum tensor \cite{Brown} to construct a divergence-free
stress energy momentum tensor. For the case of manifolds with zero curvature
boundary the finite stress energy momentum tensor is \cite{DM1}
\begin{equation}
T^{ab}=\frac{1}{8\pi }\{(\Theta ^{ab}-\Theta \gamma ^{ab})+2\alpha
(3J^{ab}-J\gamma ^{ab})-\left( \frac{n-1}{l_{\mathrm{eff}}}\right) \gamma
^{ab}\}  \label{Stres}
\end{equation}
One may note that when $\alpha $ goes to zero, the stress energy momentum
tensor (\ref{Stres}) reduces to that of Einstein gravity. To compute the
conserved charges of the spacetime, we choose a spacelike surface $\mathcal{B%
}$ in $\partial \mathcal{M}$ with metric $\sigma _{ij}$, and write the
boundary metric in ADM form:
\begin{equation}
\gamma _{ab}dx^{a}dx^{a}=-N^{2}dt^{2}+\sigma _{ij}\left( d\varphi
^{i}+V^{i}dt\right) \left( d\varphi ^{j}+V^{j}dt\right) ,
\end{equation}
where the coordinates $\varphi ^{i}$ are the angular variables
parameterizing the hypersurface of constant $r$ around the origin,
and $N$ and $V^{i}$ are the lapse and shift functions
respectively. When there is a Killing vector field $\mathcal{\xi
}$ on the boundary, then the quasilocal conserved quantities
associated with the stress energy momentum tensors of Eq.
(\ref{Stres}) can be written as
\begin{equation}
\mathcal{Q}(\mathcal{\xi )}=\int_{\mathcal{B}}d^{n-1}\varphi \sqrt{\sigma }%
T_{ab}n^{a}\mathcal{\xi }^{b},  \label{charge}
\end{equation}
where $\sigma $ is the determinant of the metric $\sigma _{ij}$, and $n^{a}$
is the timelike unit normal vector to the boundary $B$\textbf{.} For
boundaries with timelike ($\xi =\partial /\partial t$) and rotational ($%
\varsigma =\partial /\partial \varphi $) Killing vector fields, one obtains
the quasilocal mass and angular momentum
\begin{eqnarray}
M &=&\int_{\mathcal{B}}d^{n-1}\varphi \sqrt{\sigma }T_{ab}n^{a}\xi ^{b},
\label{Mas} \\
J &=&\int_{\mathcal{B}}d^{n-1}\varphi \sqrt{\sigma }T_{ab}n^{a}\varsigma
^{b},  \label{Amom}
\end{eqnarray}
provided the surface $\mathcal{B}$ contains the orbits of $\varsigma $.
These quantities are, respectively, the conserved mass and angular momentum
of the system enclosed by the boundary $\mathcal{B}$.

\section{The $(n+1)$-dimensional Charged Rotating Black Branes}

The metric of $(n+1)$-dimensional  asymptotically AdS rotating
spacetime with $k$ rotation parameters is \cite{Lemos,Awad}
\begin{eqnarray}
ds^{2} &=&-f(r)\left( \Xi dt-{{\sum_{i=1}^{k}}}a_{i}d\phi _{i}\right) ^{2}+%
\frac{r^{2}}{l^{4}}{{\sum_{i=1}^{k}}}\left( a_{i}dt-\Xi l^{2}d\phi
_{i}\right) ^{2}  \nonumber \\
&&\ \text{ }+\frac{dr^{2}}{f(r)}-\frac{r^{2}}{l^{2}}{\sum_{i<j}^{k}}%
(a_{i}d\phi _{j}-a_{j}d\phi _{i})^{2}+r^{2}dX^{2},  \label{met2}
\end{eqnarray}
where $\Xi =\sqrt{1+\sum_{i}^{k}a_{i}^{2}/l^{2}}$ and $dX^{2}$ is the
Euclidean metric on the $\left( n-1-k\right) $-dimensional submanifold. The
rotation group in $(n+1)$ dimensions is $SO(n)$ and therefore $k\leq \lbrack
n/2]$. Using the gauge potential ansatz
\begin{equation}
A_{\mu }=h(r)\left( \Xi \delta _{\mu }^{0}-\delta _{\mu }^{i}a_{i}\right)
\text{(no sum on }i\text{)}  \label{Amu}
\end{equation}
and solving Eq. (\ref{BIeq}), we obtain
\begin{equation}
h(r)=-\sqrt{\frac{n-1}{2n-4}}\frac{q}{r^{n-2}}\ {_{2}F_{1}\left( \left[
\frac{1}{2},\frac{n-2}{2n-2}\right] ,\left[ \frac{3n-4}{2n-2}\right] ,-\eta
\right) },
\end{equation}
where $q$ is an integration constant which is related to the
charge parameter, $_{2}F_{1}([a,b],[c],z)$ is hypergeometric
function and
\[
\eta =\frac{{(n-1)(n-2)q^{2}}}{2\beta ^{2}r^{2n-2}}.
\]
One may note that ${_{2}F_{1}\rightarrow 1}$ as $\eta \rightarrow
0$ ($\beta \rightarrow \infty $) and $A_{\mu }$ of Eq. (\ref{Amu})
reduces to the gauge potential of Maxwell field \cite{Deh1}. To
find the function $f(r)$ , one may use any components of Eq.
(\ref{Geq}). The simplest equation is the $rr$ component of these
equations which can be written as
\begin{eqnarray}
&&(n-1)\left[ r^{n-2}-2(n-2)(n-3)\alpha r^{n-4}\right] f^{\prime }+(n-1)(n-2)%
\left[ r^{n-3}-(n-3)(n-4)\alpha r^{n-5}f\right] f  \nonumber \\
&&+\left[ 2\Lambda +2{\beta }^{2}\left( \left( 1+\eta \right)
^{1/2}-1\right) \right] r^{n-1}=0,  \label{rreq}
\end{eqnarray}
where the prime denotes a derivative with respect to $r$. The solutions of
Eq. (\ref{rreq}) can be written as
\begin{equation}
f(r)={\frac{r^{2}}{2\,(n-2)\,(n-3)\,\alpha }}\left( 1-\sqrt{g(r)}\right) {,}
\label{f(r)}
\end{equation}
where
\begin{eqnarray}
g(r) &=&\left( 1-16\frac{{(n-3)\alpha \beta ^{2}}\eta }{n}\text{ }%
_{2}F_{1}\left( \left[ {\frac{1}{2},\frac{{n-2}}{{2n-2}}}\right] ,\left[ {%
\frac{{3n-4}}{{2n-2}}}\right] ,-{\eta }\right) \right)  \nonumber \\
&&+4\frac{(n-2)(n-3)\alpha }{n(n-1)r^{n}}\left( 2\Lambda
r^{n}+n(n-1)m-4\beta ^{2}r^{n}+4\beta ^{2}r^{n}\sqrt{1+\eta }\right) \!\!,
\label{g(r)}
\end{eqnarray}
Although the other components of the field Eq. (\ref{Geq}) are more
complicated, one can check that the metric (\ref{met2}) satisfies all the
components of Eq. (\ref{Geq}) provided $f(r)$ is given by (\ref{f(r)}, \ref
{g(r)}). Again, $f(r)$ reduces to the metric function of Ref. \cite{Deh1} as
$\beta $ goes to infinity and reduces to that of Ref. \cite{Rastegar}, when $%
\alpha \longrightarrow 0$. Using the fact that $_{2}F_{1}(a,b,c,z)$ has a
convergent series expansion for $|z|<1$, one finds that $g(r)$ for large $%
\beta $ is
\begin{eqnarray}
g(r) &=&1+\frac{8(n-2)(n-3)\alpha \Lambda }{n(n-1)}+\frac{4(n-2)(n-3)\alpha m%
}{r^{n}}-\frac{4{(n-2)}(n-3)\alpha {q^{2}}}{r^{2n-2}}  \nonumber \\
&&+\frac{{(n-1)}(n-2)^{3}(n-3)\alpha {q^{4}}}{2(3n-4)\beta ^{2}r^{4n-4}}
\label{grlarge}
\end{eqnarray}
The last term in the right hand side of the Eq. (\ref{grlarge}) is the
leading Born-Infeld correction to the Gauss-Bonnet-Maxwell black brane
solutions.

\subsection{Properties of the solutions}
One can show that the above solution is asymptotically AdS with
effective cosmological constant
\[
\Lambda _{\mathrm{eff}}=-\frac{n(n-1)}{4(n-2)(n-3)\alpha }\left( 1-\sqrt{1-%
\frac{4(n-2)(n-3)\alpha }{l^{2}}}\right)
\]
As in the case of rotating black brane solutions of Einstein-Born-Infeld
gravity, the above metric given by Eqs.(\ref{met2}, \ref{f(r)} and \ref{g(r)}%
) has an essential singularity at $r=0$, and two types of Killing and event
horizons. The Killing horizon is a null surface whose null generators are
tangent to a Killing field. It is proved that a stationary black hole event
horizon should be a Killing horizon in the four-dimensional Einstein gravity
\cite{Haw1}. This proof can not obviously be generalized to higher order
gravity, but the result is true for all the known static solutions. Although
our solution is not static, the Killing vector,
\begin{equation}
\chi =\partial _{t}+{\sum_{i}^{k}}\Omega _{i}\partial _{\phi _{i}},
\label{Kil}
\end{equation}
is the null generator of the event horizon, where $\Omega _{i}$ is the $i$th
component of angular velocity of the outer horizon which may be obtained by
analytic continuation of the metric. Setting $a_{i}\rightarrow ia_{i}$
yields the Euclidean section of (\ref{met2}), whose regularity at $r=r_{+}$
requires that we should identify $\phi _{i}\sim \phi _{i}+\beta _{+}\Omega
_{i}$. One obtains
\begin{equation}
\Omega _{i}=\frac{a_{i}}{\Xi l^{2}} \label{Om}
\end{equation}
The Hawking temperature of the black brane is
\begin{equation}
{T=}\frac{1}{2\pi }\sqrt{-\frac{1}{2}\left( \nabla _{\mu }\chi _{\nu
}\right) \left( \nabla ^{\mu }\chi ^{\nu }\right) }={{\frac{r_{+}}{2(n-1)\pi
\Xi }}}\left( 2\beta ^{2}(1-\sqrt{1+\eta _{+}})-{\Lambda }\right)
\label{Temp}
\end{equation}
Using the fact that the temperature of the extreme black brane is zero, it
is easy to show that the condition for having an extreme black hole is that
the mass parameter is equal to $m_{\mathrm{ext}}$, where $m_{\mathrm{ext}}$
is given as
\begin{eqnarray}
m_{\mathrm{ext}} &=&\frac{2(n-1)q_{\mathrm{ext}}^{n/(n-1)}}{n}\Big[\frac{%
2(2\beta ^{2}-\Lambda )}{(n-1)^{2}}\left( \frac{(n-1)(n-2)}{2\beta ^{2}\eta
_{\mathrm{ext}}}\right) ^{n/(2n-2)}+  \nonumber \\
&&{\left( \frac{(n-1)(n-2)}{2\beta ^{2}\eta _{\mathrm{ext}}}\right)
^{-(n-2)/(2n-2)}}\,_{2}F_{1}\left( \left[ \frac{1}{2},\frac{n-2}{2n-2}\right]
,\left[ \frac{3n-4}{2n-2}\right] ,-\eta _{\mathrm{ext}}\right) \Big];
\nonumber \\
\eta _{\mathrm{ext}} &=&{\frac{\Lambda \left[ \Lambda -4{\beta }^{2}\right]
}{4\beta ^{4}}}  \label{mext}
\end{eqnarray}
The metric of Eqs.(\ref{met2}, \ref{f(r)} and \ref{g(r)}) presents a black
brane solution with inner and outer horizons, provided the mass parameter $m$
is greater than $m_{\mathrm{ext}}$, an extreme black hole for $m=m_{\mathrm{%
ext}}$ and a naked singularity otherwise. Note that the horizon
radius of the extreme black brane is
\begin{equation}
r_{+\mathrm{ext}}=\left( \frac{2(n-1)(n-2)\beta ^{2}q_{\mathrm{ext}}^{2}}{%
\Lambda (\Lambda -4\beta ^{2})}\right) ^{1/(2n-2)}  \label{qext}
\end{equation}

\section{Thermodynamics \label{Therm}of black branes}
In this section we, first, calculate the thermodynamic and
conserved quantities of the black brane. Second, we obtain a
Smarr-type formula for the mass as a function of the entropy, the
angular momentum and the charge of the solution and check the
first law of thermodynamics. Finally, we perform a stability
analysis in both canonical and grand-canonical ensembles.

\subsection{Conserved and thermodynamic quantities}
Denoting the volume of the hypersurface at $r=$constant and
$t=$constant by $V_{n-1}$, the charge per unit volume of the black
brane, $Q$, can be found by calculating the flux of the
electromagnetic field at infinity, yielding
\begin{equation}
Q=\frac{(n-1)(n-2)\Xi }{8\pi }q.  \label{Charg}
\end{equation}
The electric potential $\Phi $, measured at infinity with respect to the
horizon, is defined by \cite{Gub}
\begin{equation}
\Phi =A_{\mu }\chi ^{\mu }\left| _{r\rightarrow \infty }-A_{\mu }\chi ^{\mu
}\right| _{r=r_{+}},  \label{Pot}
\end{equation}
where $\chi $ is the null generator of the horizon given by Eq.
(\ref{Kil}). One finds
\begin{equation}
\Phi =\sqrt{\frac{(n-1)}{2(n-2)}}\frac{q}{\Xi r_{+}^{n-2}}\text{ }%
_{2}F_{1}\left( \left[ {\frac{1}{2},\frac{{n-2}}{{2n-2}}}\right] ,\left[ {%
\frac{{3n-4}}{{2n-2}}}\right] ,-{\eta }\right) .
\end{equation}
Using Eqs. (\ref{Mas}) and (\ref{Amom}), the mass and angular
momenta per unit volume $V_{n-1}$ of the solution are calculated
as
\begin{eqnarray}
M &=&\frac{1}{16\pi }m\left( n\Xi ^{2}-1\right) ,  \label{Mass} \\
J_{i} &=&\frac{1}{16\pi }n\Xi ma_{i}.  \label{Angmom}
\end{eqnarray}
Black hole entropy typically satisfies the so-called area law,
which states that the entropy of a black hole equals one-quarter
of the area of its horizon \cite{Beck}. This near universal law
applies to almost all kinds of black holes and black strings in
Einstein gravity \cite{Haw2}. However in higher derivative gravity
the area law is not satisfied in general \cite {fails}. For
asymptotically flat black hole solutions of \cite{Myers2}
\begin{equation}
S=\frac{1}{4}\sum_{k=1}^{[(d-1)/2]}k\alpha _{k}\int d^{n-1}x\sqrt{\tilde{g}}%
\tilde{\mathcal{L}}_{k-1}  \label{Enta}
\end{equation}
where the integration is done on the $(n-1)$-dimensional spacelike
hypersurface of Killing horizon, $\tilde{g}_{\mu \nu }$ is the induced
metric on it, $\tilde{g}$ is the determinant of $\tilde{g}_{\mu \nu }$ and $%
\tilde{\mathcal{L}}_{k}$ is the $k$th order Lovelock Lagrangian of $\tilde{g}%
_{\mu \nu }$. The asymptotic behavior of the topological class of black
holes we are considering is AdS, and therefore we calculate the entropy
through the use of Gibbs-Duhem relation
\begin{equation}
S=\frac{1}{T}(\mathcal{M}-\Gamma _{i}\mathcal{C}_{i})-I  \label{GibsDuh}
\end{equation}
where $I$ is the finite total action evaluated on the classical
solution, and $\mathcal{C}_{i}$ and $\Gamma _{i}$ are the
conserved charges and their associate chemical potentials
respectively. Using eqs. (\ref{IG}) and (\ref {Ict}), the finite
action per unit volume $V_{n-1}$ can be calculated as
\begin{equation}
I=-\frac{\Xi }{4n}\left(
r_{+}^{n-1}+\frac{2(n-1)^{2}l^{2}q^{2}\text{
}_{2}F_{1}\left( \left[ {\frac{1}{2},\frac{{n-2}}{{2n-2}}}\right] ,\left[ {%
\frac{{3n-4}}{{2n-2}}}\right] ,-{\eta }_{+}\right) }{r_{+}^{n-1}[4\beta
^{2}l^{2}(1-\sqrt{1+\eta _{+}})+n(n-1)]}\right)  \label{finiteACT}
\end{equation}
Now using Gibbs-Duhem relation (\ref{GibsDuh}) and eqs. (\ref{Charg}) - (\ref
{Angmom}) and (\ref{finiteACT}) one obtains
\begin{equation}
S=\frac{\Xi}{4}r_{+}^{(n-1)}.  \label{Entropy}
\end{equation}
for the entropy per unit volume $V_{n-1}$. This shows that the entropy obeys
the area law for our case where the horizon curvature is zero.

\subsection{Energy as a function of entropy, angular momenta and charge}

Calculating all the thermodynamic and conserved quantities of the black
brane solutions, we now check the first law of thermodynamics for our
solutions. We obtain the mass as a function of the extensive quantities $S$,
$\mathbf{J}$, and $Q$. Using the expression for the entropy, the mass, the
angular momenta, and the charge given in Eqs. (\ref{Entropy}), (\ref{Charg}%
), (\ref{Mass}), (\ref{Angmom}), and the fact that $f(r_{+})=0$, one can
obtain a Smarr-type formula as
\begin{equation}
M(S,\mathbf{J},Q)=\frac{(nZ-1)J}{nl\sqrt{Z(Z-1)}},  \label{Smar}
\end{equation}
where $J=\left| \mathbf{J}\right| =\sqrt{\sum_{i}^{k}J_{i}^{2}}$ and $Z=\Xi
^{2}$ is the positive real root of the following equation:

\begin{eqnarray}
&&\left( n(n-1)+4\beta ^{2}l^{2}\right) S^{n/(n-1)}-4\beta l^{2}S^{1/(n-1)}%
\sqrt{\pi ^{2}Q^{2}+\beta ^{2}S^{2}}-\frac{4^{(n-2)/(n-1)}(n-1)\pi
lJZ^{1/(2n-2)}}{\sqrt{(Z-1)}}  \nonumber \\
&&+4\frac{n-1}{n-2}\pi ^{2}l^{2}Q^{2}S^{-(n-2)/(n-1)}\ {_{2}F_{1}\left( %
\left[\frac{1}{2},\frac{n-2}{2n-2}\right],\left[\frac{3n-4}{2n-2}\right],-{%
\frac{\pi ^{2}Q^{2}}{\beta ^{2}S^{2}}}\right)} =0.  \label{Zsmar}
\end{eqnarray}
One may then regard the parameters $S$, $J_{i}$'s, and $Q$ as a complete set
of extensive parameters for the mass $M(S,\mathbf{J},Q)$ and define the
intensive parameters conjugate to them. These quantities are the
temperature, the angular velocities, and the electric potential
\begin{equation}
T=\left( \frac{\partial M}{\partial S}\right) _{J,Q},\ \ \Omega _{i}=\left(
\frac{\partial M}{\partial J_{i}}\right) _{S,Q},\ \ \Phi =\left( \frac{%
\partial M}{\partial Q}\right) _{S,J}.  \label{Dsmar}
\end{equation}
It is a matter of straightforward calculation to show that the intensive
quantities calculated by Eq. (\ref{Dsmar}) coincide with Eqs. (\ref{Om}), (%
\ref{Temp}), and (\ref{Pot}). Thus, these quantities satisfy the first law
of thermodynamics
\[
dM=TdS+{{{\sum_{i=1}^{k}}}}\Omega _{i}dJ_{i}+\Phi dQ.
\]

\subsection{Stability in the canonical and the grand-canonical Ensemble}
\begin{figure}[tbp]
\epsfxsize=10cm \centerline{\epsffile{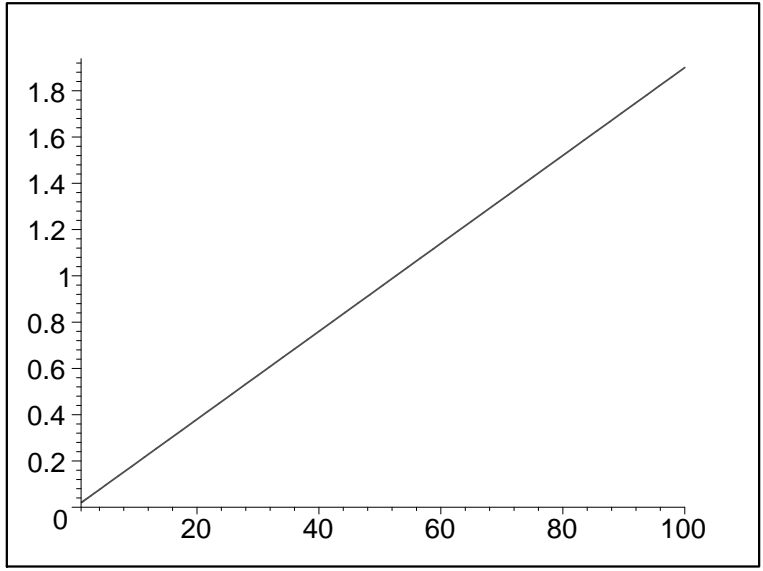}}
\caption{$C_{\mathbf{J},Q}$ versus $\Xi $ for $n=6$, $l=1$, $\protect\beta %
=1 $, $q=0.7$ and $r=0.8$.} \label{Fig1}
\end{figure}

Finally, we investigate the stability of charged rotating black brane
solutions of Born-Infeld gravity. The stability of a thermodynamic system
with respect to the small variations of the thermodynamic coordinates, is
usually performed by analyzing the behavior of the entropy $S(M,Q,\mathbf{J}%
) $ near equilibrium. The local stability in any ensemble requires that $%
S(M,Q,\mathbf{J})$ be a concave function of its extensive variables or that
its Legendre transformation is a convex function of the intensive variables%
\textbf{.} The stability can also be studied by the behavior of the energy $%
M(S,Q,\mathbf{J})$ which should be a convex function of its extensive
variable. Thus, the local stability can in principle be carried out by
finding the determinant of the Hessian matrix of $M(S,Q,\mathbf{J})$ with
respect to its extensive variables $X_{i}$, $\mathbf{H}_{X_{i}X_{j}}^{M}=[%
\partial ^{2}M/\partial X_{i}\partial X_{j}]$ \cite{Gub}. In our case the
entropy $S$ is a function of the mass, the angular momenta, and the charge.
The number of thermodynamic variables depends on the ensemble that is used.
In the canonical ensemble, the charge and the angular momenta are fixed
parameters, and therefore the positivity of the heat capacity $C_{\mathbf{J}%
,Q}=T_{+}/(\partial ^{2}M/\partial S^{2})_{\mathbf{J},Q}$ is sufficient to
ensure local stability. $(\partial ^{2}M/\partial S^{2})_{\mathbf{J},Q}$ at
constant charge and angular momenta is
\begin{eqnarray}
\left( \frac{\partial ^{2}M}{\partial S^{2}}\right) _{_{\mathbf{J},Q}}
&=&\Upsilon ^{-1}\Bigg\{\frac{\left( n-1\right) l^{2}m}{r_{+}^{n}}\left[
(n-2)\Xi ^{2}+1\right] \times \Bigg.  \nonumber \\
&&\left\{ n(n-1)\sqrt{1+\eta _{+}}+4\left( n-2\right) l^{2}{\beta }^{2}\eta
_{+}+4l^{2}\beta ^{2}\left( \sqrt{1+\eta _{+}}-1\right) \right\} +\Bigg.
\nonumber \\
&&n\Big\{32l^{4}\beta ^{4}\left( \sqrt{1+\eta _{+}}-1\right) +n\left(
n-1\right) \left( n(n-1)\sqrt{1+\eta _{+}}-8l^{2}{\beta }^{2}\eta
_{+}\right) +\Big.\Bigg.  \nonumber \\
&&8nl^{2}\beta ^{2}\left[ 2l^{2}{\beta }^{2}\eta _{+}\left( \sqrt{1+\eta _{+}%
}-2\right) +n(n-1)\left( \sqrt{1+\eta _{+}}-1\right) \right] \Big\}(\Xi
^{2}-1)\Bigg\};  \nonumber \\
\Upsilon &=&(n-1)^{3}\pi m\Xi ^{2}l^{4}\left[ (n-2)\Xi ^{2}+1\right] \left(
1+\eta _{+}\right) ^{1/2}r_{+}^{2}  \label{dMSS}
\end{eqnarray}
The heat capacity is positive for $m\geq m_{\mathrm{ext}}$, where the
temperature is positive. This fact can be seen easily for $\Xi =1$, where
the second term of Eq. (\ref{dMSS}) is zero and the first term is positive.
Also, one may see from Fig. \ref{Fig1} that the heat capacity increases as $%
\Xi $ increases, and therefore it is always positive. Thus, the black brane
is stable in the canonical ensemble. In the grand-canonical ensemble, after
some algebraic manipulation, we obtain
\begin{equation}
H_{S\mathbf{J}Q}^{M}=\frac{\Pi }{\Psi }
\end{equation}
where
\begin{eqnarray}
\Pi &=&\frac{64\pi }{l^{2}\Xi ^{6}[(n-2)\Xi ^{2}+1]}\Bigg\{\left( 2\beta
^{2}(\left( 1+\eta _{+}\right) ^{1/2}-1)+(n-1)(n-2)^{2}q^{2}r_{+}^{-2n+2}{%
-\Lambda }\left( 1+\eta _{+}\right) ^{1/2}\right) \Bigg.  \nonumber \\
&&\times {_{2}F_{1}\left( \left[ \frac{1}{2},\frac{n-2}{2n-2}\right] ,\left[
\frac{3n-4}{2n-2}\right] ,-\eta _{+}\right) }+(n-2)\newline
\left( (2\beta ^{2}\newline
-\Lambda ){-2}\beta ^{2}\left( 1+\eta _{+}\right) ^{1/2}\right) \Bigg\}
\end{eqnarray}
and
\begin{eqnarray}
\Psi &=&(n-1)(n-2)r_{+}^{n-2}\left( 1+\eta _{+}\right) ^{1/2}\Bigg\{%
(n-1)^{2}q^{2}{_{2}F_{1}\left( \left[ \frac{1}{2},\frac{n-2}{2n-2}\right] ,%
\left[ \frac{3n-4}{2n-2}\right] ,-\eta _{+}\right) }\Bigg.  \nonumber \\
&&+r_{+}^{2n-2}\left( (2\beta ^{2}-\Lambda )-2\beta ^{2}\left( 1+\eta
_{+}\right) ^{1/2}\right) \Bigg\},
\end{eqnarray}

Since the value of $_{2}F_{1}\left( \left[ \frac{1}{2},\frac{n-2}{2n-2}%
\right] ,\left[ \frac{3n-4}{2n-2}\right] ,-\eta _{+}\right) $ is between $0$
and $1$ and $2\beta ^{2}\left( 1+\eta _{+}\right) ^{1/2}\leq (2\beta
^{2}-\Lambda )$ for $q\leq q_{\mathrm{ext}}$, it is easy to see that $H_{S%
\mathbf{J}Q}^{M}$ is positive for all the allowed values of
\begin{equation}
q\leq q_{\mathrm{ext}}=\frac{r_{+}^{n-1}}{\beta }\sqrt{\frac{\Lambda
(\Lambda -4\beta ^{2})}{2(n-1)(n-2)}}  \label{qless}
\end{equation}
Thus, the $(n+1)$-dimensional asymptotically AdS charged rotating
black brane is locally stable in the grand-canonical ensemble.

\section{ CLOSING REMARKS}

In this paper, we found a new class of rotating solutions in
Gauss-Bonnet-Born-Infeld gravity in the presence of the
cosmological constant. These solutions which are asymptotically
AdS and have flat horizon may be interpreted as black brane
solutions with inner and outer event horizons provided the mass
parameter $m$ is greater than an extremal value given by
Eq.(\ref{mext}), an extreme black brane for $m=m_{\mathrm{ext}}$
and a naked singularity otherwise. We found that these solutions
reduce to the solutions of Eistein-Born-Infeld gravity as $\alpha
\longrightarrow 0,$ and reduce to those of Gauss--Bonnet-Maxwell
gravity as $\beta \longrightarrow \infty $. The counterterm method
inspired by the AdS/CFT correspondence has been widely applied to
the case of Einstein gravity. Here, we applied this method to the
solutions of Gauss-Bonnet-Born-Infeld
gravity with flat boubdary at $r=$\textrm{constant }and\textrm{\ }$t=$%
\textrm{constant}, and calculated the finite action and conserved quantities
of them. The physical properties of the brane such as the temperature, the
angular velocity, the electric charge and the potential have been computed.
We found that the conserved quantities of the black brane do not depend on
the Gauss-Bonnet parameter $\alpha $. Then, we obtained the entropy of the
black brane through the use of Gibbs-Duhem relation and found that it obeys
the area law of entropy. We also obtained a Smarr-type formula for the mass
as a function of the extensive parameters $S$, $\mathbf{J}$\thinspace\ and $Q
$, calculated the intensive parameters, temperature, angular velocity, and
electric potential, and showed that these quantities satisfy the first law
of thermodynamics. We also studied the phase behavior of the $(n+1)$%
-dimensional rotating black branes in GBBI gravity and showed that
there is no Hawking-Page phase transition in spite of the presence
of a nonlinear electromagnetic field. Indeed, we calculated the
heat capacity and the
determinant of the Hessian matrix of the mass with respect to $S$, $\mathbf{J%
}$ and $Q$ of the black branes and found that they are positive
for all the phase space, which means that the brane is stable for
all the allowed values of the metric parameters discussed in Sec.
\ref{Therm}. This phase behavior is commensurate with the fact
that there is no Hawking-Page transition for a black object whose
horizon is diffeomorphic to $\Bbb{R}^{p}$ and therefore the system
is always in the high temperature phase \cite{Wit2}. Although, the
presence of the nonlinear Born-Infeld field has no effects on the
stability of the black brane, the presence of a scalar field makes
the solution unstable \cite{SDRP}.

\acknowledgments{This work has been supported by
Research Institute for Astronomy and Astrophysics of Maragha,
Iran}

\end{document}